\documentclass{camera}
\usepackage{epsfig}

\def\cO#1{{\cal O}\left( {#1} \right)}

\def\as{\alpha_{\mathsf{s}}}

\def\cO#1{{\cal{O}}\left(#1\right)}
\def\al{\alpha}
\def\as{\alpha_{\mbox{\scriptsize s}}}

\def\be{\beta}

\def\PT{\mbox{\scriptsize PT}}
\def\NP{\mbox{\scriptsize NP}}

\def\ee{e^+e^-}

\def\cM{{\cal{M}}}

\def\eps{\epsilon}

\def\LQCD{\Lambda_{\mbox\scriptsize QCD}}

\def\VEV#1{\left\langle#1\right\rangle}

 \newskip\humongous \humongous=0pt plus 1000pt minus 1000pt
   \newif\ifdtup

\def\fun#1#2{\lower3.6pt\vbox{\baselineskip0pt\lineskip.9pt
  \ialign{$\mathsurround=0pt#1\hfil##\hfil$\crcr#2\crcr\sim\crcr}}}

\def\Journal#1#2#3#4{#1{\bf #2} (#4) #3}
\def\NPB{{\em Nucl.\ Phys.} B}

\def\PLB{{\em Phys.\ Lett.}  B}

\def\ZPC{{\em Z.\ Phys.} C}
\def\JHEP{{\em JHEP }}

\def\PREP{{\em Phys.\ Rep. }}
\def\NPPS{{\em  Nucl.\ Phys.\ Proc.\ Suppl.\ }}

\begin{document}

%
\title{NON-PERTURBATIVE EFFECTS IN $e^+e^-$ EVENT-SHAPE VARIABLES}

%
\author{A. Banfi$^{1}$ \And G. Zanderighi$^{2}$}

%
\organization{ $^1$Dipartimento di Fisica, Universit\`a di Milano-Bicocca and
 INFN, Sezione di Milano, Italy;\\ 
$^2$Dipartimento di Fisica Nucleare
 e Teorica, Universit\`a di Pavia and INFN, Sezione di Pavia, Italy.
 }

\maketitle

\abstract{We review theoretical methods employed to study non-perturbative
 contributions to $\ee$ event-shapes and discuss their
 phenomenological relevance.}

%
\section{Introduction}
Event-shape variables in $\ee$ annihilation, such as Thrust $T$,
Heavy-jet Mass $M_H$, $C$-parameter, Broadening $B$, have provided
various tests of QCD and a way to measure $\alpha_s$.  Besides the
perturbative (PT) results, agreement with data is achieved only by
taking into account additional corrections of non-perturbative (NP)
origin recently addressed through analytic approaches [1--3].

As explained in Section 2, the net effect of these corrections is to
raise the mean value of an event-shape by an amount proportional to
$1/Q$, being $Q$ the center of mass energy.  Similar features occur
for distributions. In Section 3 {\it universality} of $1/Q$
corrections is discussed and it is mentioned how hadron mass effects
partially spoil
the 'simple' universality picture. 
We conclude in Section 4 giving some outlooks.

\section{Power correction to event shapes}
\begin{figure}[ht]
\begin{center}
\epsfig{file=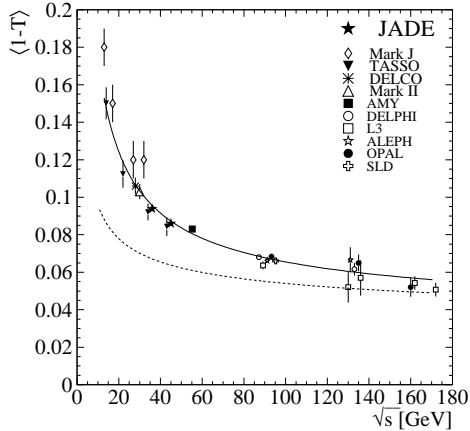,width=0.4\textwidth}
\caption{Mean value of $1-T$ as a function of $Q$ \cite{JADE}.
 The dotted line represents the NLO prediction, the solid line the
 improved prediction including a $1/Q$ correction.
\label{fig:meanT}}
\end{center}
\end{figure}

To see how NP corrections to event-shapes emerge, 
we consider, for instance, the mean value of $\tau\equiv 1-T$. 
As shown in fig.~\ref{fig:meanT}, a 
next-to-leading order
(NLO)
calculation alone does not fully describe data and one has
to add a power suppressed contribution of the form $C_{\tau}/Q$, with
$C_{\tau}\simeq 1\mbox{GeV}$.

One cannot get rid of this mismatch by simply taking into account
higher orders in the PT expansion, since the PT series itself is
divergent (for a recent review see \cite{renormalons}). Any attempt to
give a meaning to the series leads to an ambiguity (infra-red
renormalon) which, for event shapes, is of the order $\LQCD/Q$. This
ambiguity must be canceled by a NP contribution with the same power
behavior.

For the distribution the situation is more involved. Actually, the
full $\tau$ distribution can be expressed as a convolution of a PT
distribution with a NP shape function $f_{\NP}$. In the region
$\tau\gg\LQCD/Q$ one can make the approximation
\begin{equation}
\label{eq:distT}
\frac{d\sigma}{\sigma d\tau}=\int d\eps f_{\NP}(\eps)
\frac{d\sigma_{\PT}(\tau-\frac{\eps}{Q})}{\sigma d\tau}
\simeq
\frac{d\sigma_{\PT}(\tau-\frac{\VEV \eps}{Q})}{\sigma d\tau}\>, 
\end{equation}
so that hadronization corrections result in a $1/Q$ power-suppressed
shift of the PT distribution.
In the region $\tau\sim\LQCD/Q$ higher powers become equally
important, so that the full shape function should be
kept. This is the basis of the Korchemsky-Sterman approach
\cite{KS}, where a parameterization of the shape
function is given and the parameters are fitted to the data.

\section{Universality of $1/Q$ power corrections}
Although $1/Q$ power corrections are intrinsically NP quantities, one
is able to predict their relative size from one observable to the next.
First one observes that
hadron multiplicity $ n_h$ is uniform in rapidity ($\eta$):
\begin{equation}
\label{eq:multi}
\frac{d n_h}{d\ln k_t d\eta}=\phi_h(k_t)\>,
\end{equation}
with $k_t$ the particle transverse momentum with respect to the thrust
axis. This is the basis of the local parton hadron duality (LPHD)
approach~\cite{LPHD}, where hadron momentum flow is supposed to follow
parton flow. Furthermore, a soft particle contribution to an event
shape may be written as the product of $k_t/Q$ times an
observable-dependent function of rapidity $f_V(\eta)$ (for the Thrust
$f_{\tau}(\eta)=e^{-|\eta|}$).  As a consequence, the NP correction to
the mean value of $V$ can be expressed as (see for
example~\cite{Milan}, and references therein)
\begin{equation}
\label{eq:meanV}
\VEV{V}_{\NP}\!=\!\frac{\VEV{k_t}_{\NP}}{Q}\>c_V\>,\quad
\>\mbox{with}\quad 
\frac{\VEV{k_t}_{\NP}}{Q}\!=\!
\int \frac{dk_t}{k_t}\frac{k_t}{Q}\sum_h \phi_h(k_t)\>,\quad
c_V\!=\!\int d\eta f_V(\eta)\>, 
\end{equation}
so that all the observable dependence is contained in the calculable
coefficient $c_V$ and one is left with  only one unknown NP parameter
$\VEV{k_t}_{\NP}$.  This property of $1/Q$ power corrections is
commonly referred to as {\it universality}.

\subsection{Dispersive method}
\begin{figure}[ht]
\begin{center}
\epsfig{file=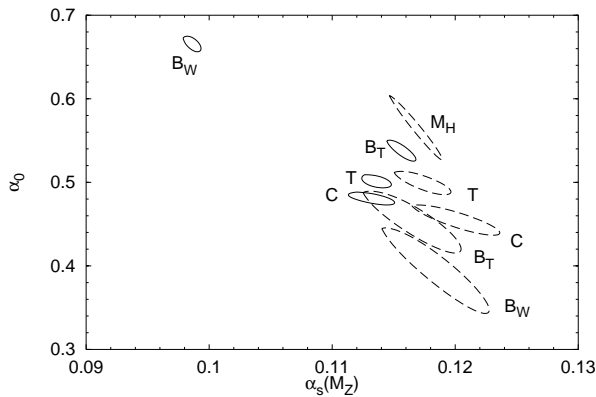
,width=0.5\textwidth}	
\caption{$2$-$\sigma$ contours for fits to various two-jet 
observables~\cite{SZ}.  Solid curves indicate fits to distributions,
while dashed lines indicate fits to mean values.
\label{fig:as&a0}}
\end{center} 
\end{figure}

A useful parameterization of $\VEV{k_t}_{\NP}$ is provided by the
dispersive approach \cite{DMW}, in which the running coupling is
defined at any scale through a dispersion relation. The NP parameter
$\VEV{k_t}_{\NP}$ is related to $\al_0$, the average of the dispersive
coupling below a certain low scale $\mu_I$ (conventionally chosen to
be $\mu_I=2\>\mbox{GeV}$):
\begin{equation}
\label{eq:alpha0}
 \frac{\VEV{k_t}_{\NP}}{Q}=\frac{4C_F}{\pi^2}\cM\frac{\mu_I}{Q}
\al_0(\mu_I)+\cO{\as(Q)\frac{\mu_I}{Q}}\>,\qquad 
\al_0(\mu_I)=\frac{1}{\mu_I}\int_0^{\mu_I}\!dk\>\as(k)\>.
\end{equation}
Here the Milan factor $\cM$ accounts for the non-inclusiveness of
shape variables~\cite{Milan}.  The value of $\al_0$ has been measured
by performing a simultaneous fit of $\as$ and $\al_0$ to both mean
values and distributions, and is found to be consistent with the
universality hypothesis\footnote{Actually, the $1/Q$ corrections to
the wide-jet broadening $B_W$ distribution are  not yet believed to be
fully understood. This seems to be a common problem of less inclusive
quantities, such as $B_W$ and $M_H$, where one particular hemisphere
is chosen.}, as shown in fig.~\ref{fig:as&a0}.

These fits have been performed only with two-jet event shapes.  Only
very recently these studies have been extended to three-jet events,
and the power corrections to the thrust-minor and $D$-parameter have
been addressed \cite{BDMZ}. 

\subsection{Mass effects and universality}
Until now hadron masses have been neglected in the definition of the
event-shapes. As shown in \cite{masses}, hadron masses give rise to
additional power corrections $\delta V_m$ which, in general,  
are not proportional to $c_V$, thus spoiling the universality
picture. However, with a suitable redefinition of the observables
(E-scheme) one is able to eliminate the non-universal contributions
leaving just universal mass corrections of the form
\begin{equation}
\label{eq:mass}
\delta V_m = c_V\frac{\mu_{\ell}}{Q}\ln^A\frac{\LQCD}{Q}\>,
\qquad A=4C_A/\be_0\simeq 1.6\>,	
\end{equation}
with $\mu_{\ell}$ a new unknown parameter which depends on the hadron
level considered.  Unfortunately, currently available data are not
precise enough to extract $\al_0$ and $\mu_{\ell}$
simultaneously. However, changing the definition scheme
or the hadron level results in systematic uncertainties in $\as$ and
$\al_0$ fits, thus revealing the presence of mass effects of the form
predicted by eq.~\ref{eq:mass}.

\section{Conclusions}
During this decade much theoretical and experimental effort has been
devoted to the study of hadronization effects in $e^+e^-$ event
shapes.  Experiments have confirmed the universality of $1/Q$ power
corrections, thus supporting the validity of the LPHD
approach. However, more refined analyses have revealed the need to
include higher power corrections through a NP shape function and the
existence of mass effects.  These and related topics need further
experimental investigation.

\paragraph{Acknowledgements}
We are thankful to Matteo Cacciari, Pino Marchesini, Gavin Salam and
Graham Smye for helpful comments and suggestions, and to the organizers
of LEPTRE conference for the pleasant days we spent in Rome.

%
\end{document}